\documentclass[aps,prl,reprint,twocolumn,amsmath,amssymb,citeautoscript,longbibliography]{revtex4-1}
\usepackage{graphicx}
\usepackage{appendix}
\usepackage{csquotes}
\usepackage{dcolumn}
\usepackage{bm}
\usepackage{latexsym,epsfig}
\usepackage{graphicx}
\usepackage{verbatim}
\usepackage{comment}
\usepackage{amsmath}
\usepackage{amssymb}
\usepackage{physics}
\usepackage{stmaryrd}
\usepackage{color}
\usepackage{epstopdf}
\usepackage{grffile}
\usepackage{lipsum}
\usepackage{enumitem}
\usepackage{ulem}
\usepackage{bbold}

\usepackage[dvipsnames]{xcolor}
\DeclareGraphicsExtensions{.eps}

\newcommand{\beq}{\begin{equation}}
\newcommand{\eeq}{\end{equation}}
\newcommand{\bea}{\begin{eqnarray}}
\newcommand{\eea}{\end{eqnarray}}
\newcommand{\ben}{\begin{eqnarray*}}
\newcommand{\een}{\end{eqnarray*}}
\newcommand{\bfig}{\begin{figure}}
\newcommand{\efig}{\end{figure}}

\usepackage{hyperref}
\hypersetup{
    colorlinks=true,      
    urlcolor=blue,
    citecolor=blue,
    linkcolor=blue
}
\begin{document}
\title{Reentrant topology and reverse pumping in a quasiperiodic flux ladder} 
\author{Sanchayan Banerjee}
\affiliation{School of Physical Sciences, National Institute of Science Education and Research, Jatni 752050, India}
\affiliation{Homi Bhabha National Institute, Training School Complex, Anushaktinagar, Mumbai 400094, India}

\author{Rajashri Parida}
\affiliation{School of Physical Sciences, National Institute of Science Education and Research, Jatni 752050, India}
\affiliation{Homi Bhabha National Institute, Training School Complex, Anushaktinagar, Mumbai 400094, India}

\author{Tapan Mishra}
\email{mishratapan@niser.com}
\affiliation{School of Physical Sciences, National Institute of Science Education and Research, Jatni 752050, India}
\affiliation{Homi Bhabha National Institute, Training School Complex, Anushaktinagar, Mumbai 400094, India}

\date{\today}

\begin{abstract}
Topological phases of matter are known to be unstable against strong onsite disorder in one dimension. In this work, however, we propose that in the case of a topological ladder, an onsite quasiperiodic disorder under proper conditions, first destroys the initial topological phase and subsequently, induces another topological phase through a gap-closing point. Remarkably, by allowing a staggered flux piercing through the plaquettes of the ladder, the gapless point bifurcates into two gapless critical lines, resulting in a trivial gapped phase sandwiched between the two topological phases. This results in a scenario where the system first undergoes a transition from one topological phase to a trivial phase and then to the other topological phase as a function of the quasiperiodic disorder strength. Such disorder induced re-entrant topological phase transition reveals a phenomenon of direction reversal in the topological transport, which we identify through Thouless charge pumping. 

\end{abstract}
\maketitle
\paragraph*{Introduction.-} 
\label{sec:intro}
The topological insulators are hallmarked by the presence of degenerate edge states in the bulk gap, associated with the bulk topological invariant, i.e., the bulk-boundary correspondence. These phases carry enormous fundamental significance and are very promising for technological applications due to their robustness against various perturbations~\cite{senthil_review_2015,rachel_review, vonKlitzing2017,Fidkowski,Oshikawa,Xiao-Gang,sptinteract,Pachos_2014,extended_ssh,glide_ssh_ladder,chiral_dynamics}. In particular, the topological phases exhibit exceptional stability when the system is subjected to finite disorder.
However, stronger disorder poses a serious threat and often leads to the breakdown of the topological nature due to the closing up of the bulk gap~\cite{hayward_rice_mele, Nakajima_disorder, topo_anderson_pump, loc_topology_r_citro, disorder_pump_hoti,pumping_photonics}. This understanding was countered with the discovery of the topological Anderson insulators, where onsite disorder is known to promote transitions to topological insulators from either a metal or a trivial insulator in two-dimensional systems~\cite{TAI1, TAI2,TAI3}. However, such a remarkable phenomenon in one-dimensional models does not occur in systems with onsite disorder due to the breaking up of the symmetry that is necessary to favor a topological phase, although exceptions are found in one-dimensional lattices with hopping disorder due to the preservation of the required symmetry~\cite{BRYCETAI, 1dTAI1}.

The question remained open whether onsite disorder of any type (random or quasiperiodic) in one dimension can favor a topological to trivial phase transition or vice-versa through a gap closing point or not. In this context, a recent study by some of us has shown that in the presence of dimerized quasiperiodic onsite disorder,  the gap in the topological phase first closes and reopens again, leading to a topological to trivial phase transition~\cite{ashirbad_disorder_pumping}. In other words, although increasing disorder favors a reopening up of the bulk gap, it does not preserve the topological nature of the phase. At this point, it is important to ask if an onsite disorder can really favor a topological phase instead of destroying it. 

\begin{figure}[t]
\begin{center}
\includegraphics[width=0.8\columnwidth]{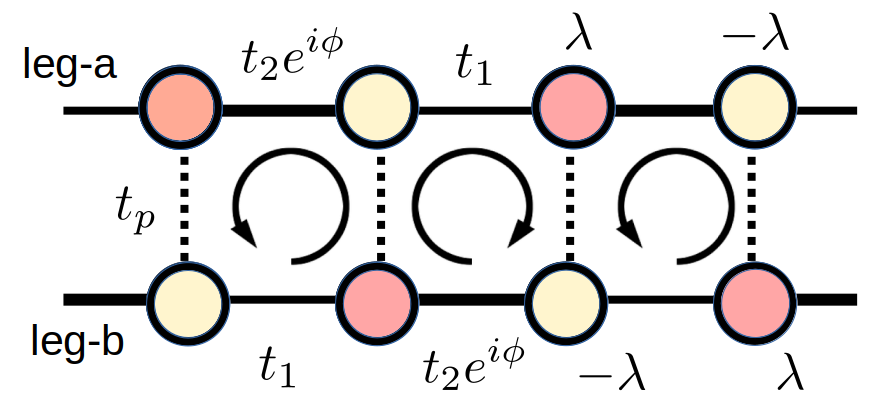}
\end{center}
\caption{
Schematic representation of the model. 
The rung hopping $ t_p $ is depicted by dotted lines, while intra-chain hoppings are represented by thin solid lines ($ t_1 $) and thick solid lines ($ t_2 e^{i\phi} $), indicating alternating dimerization along the upper and lower legs of the ladder. }
\label{figure 1}
\end{figure}

In this study, we address this question by proposing a model where the interplay of onsite quasiperiodic disorder and magnetic field in a two-leg ladder, as shown in Fig.~\ref{figure 1}, reveals that under proper conditions, disorder can induce topological phase transitions.
By starting from a topological phase, we demonstrate that the quasiperiodic disorder first destroys the topological phase by closing the bulk gap at a critical disorder strength, as expected. Interestingly, however, a further increase in the strength of disorder results in a reopening of the gap and appearance of another topological phase. We obtain that the new topological phase possesses topological characters that are opposite to those of the former. This makes the two topological phases distinctly different from each other. Surprisingly, with the onset of a finite flux in the system, the gap-closing point bifurcates into two gapless lines that separate the two topological phases from a gapped trivial phase in between up to certain values of flux strengths. This results in a remarkable onsite disorder induced re-entrant topological phase transition where a topological phase first undergoes a transition to a trivial phase and then to a topological phase of different character. Such re-entrant topological phase transition leads to a reversal of quantized particle transport with the modulation of disorder strength, which is revealed by the Thouless charge pumping mechanism.

\paragraph*{Model.-}
\label{sec:model}

The Hamiltonian of the system consisting of spinless fermions is given by
\begin{equation}
    H = H_0 + H_D.
    \label{eq:ham}
\end{equation}
$H_0$ is the kinetic term, which is  given by:
\begin{equation}
\begin{split}
H_0 &= -\sum_j \Big[ t_p (a_j^\dagger b_j + \text{H.c.}) \Big]\\
&- \sum_j \Big[ f(j) (a_j^\dagger a_{j+1} + \text{H.c.}) + g(j) (b_j^\dagger b_{j+1} + \text{H.c.}) \Big].
\label{eq:kin_part}
\end{split}
\end{equation}
Here, the functions $ f(j) $ and $ g(j) $ define the hopping amplitudes in the upper and lower legs, respectively, and are defined as:
\begin{equation}
\begin{split}
f(j) &= t_1 \delta_{j \bmod 2, 1} + t_2 e^{i\phi}  \delta_{j \bmod 2, 0}, \\
g(j) &= t_1 \delta_{j \bmod 2, 0} + t_2 e^{i\phi} \delta_{j \bmod 2, 1}.
\end{split}
\end{equation}

The operators $a_j^\dagger$ ($b_j^\dagger$) and $a_j$ ($b_j$) represent the creation and annihilation operators for spinless fermions on leg-a (b), respectively, where $j$ represents the rung index. $ t_1 $ and $ t_2 $ represent the dimerized hopping amplitudes along the upper and lower legs of the ladder, respectively. The phase factor $ \phi $ associated with $t_2$ introduces an effective magnetic flux through the plaquettes of the ladder. The presence of a magnetic flux renders the $t_2$ hopping term complex, introducing a phase factor $\phi = \frac{e}{\hbar} \int_{r_i}^{r_f} A(r)\cdot \mathrm{d}r$, where $A(r)$ denotes the magnetic vector potential~\cite{gauge_potential_rev}. In our calculations, we express the phase as $\phi = 2\pi \Phi/\Phi_0$, with $\Phi$ representing the magnetic flux through each plaquette and $\Phi_0$ is the magnetic flux quantum. $t_p$ denotes the hopping amplitude along the rungs. 
The introduction of flux in this system under consideration breaks both particle-hole and time-reversal symmetries, while preserving chiral symmetry~\cite{10fold1, Ryu_2010} in this clean limit.
The second term $H_D$ is the disorder term which is given by:
$H_D = \sum_j \Big[ h_A(j) n_{a,j} + h_B(j) n_{b,j} \Big]$,
where the onsite potentials $ h_A(j) $ and $ h_B(j) $ follow a staggered quasiperiodic pattern within each unit cell such that :
\begin{equation}
\begin{split}
h_A(j) &= (-1)^j \lambda \cos\big(2\pi\beta (2j + j \bmod 2) +\theta\big), \\
h_B(j) &= (-1)^{j+1} \lambda \cos\big(2\pi\beta (2j + (j+1) \bmod 2)+\theta\big).
\end{split}
\end{equation}
Here,  $n_{a,j} = a_j^\dagger a_j, \quad n_{b,j} = b_j^\dagger b_j$ are the onsite number operators on the legs of the ladder distinguished by the leg-indices. 
The quasiperiodicity is ensured by choosing $\beta = (\sqrt{5}-1)/2$ (the inverse golden ratio), with the disorder strength tuned by $\lambda$, and $\theta$ is the phase in the onsite potential.
In our calculations, we set $ t_1 = 1 $ as the energy scale and fix the system size (total number of sites) to $ L = 2584$ ($1292$ rungs). This setup allows us to explore the interplay of dimerization, flux, and quasiperiodic disorder in the system.

\begin{figure}[t]
\centering
\includegraphics[width=1\columnwidth]{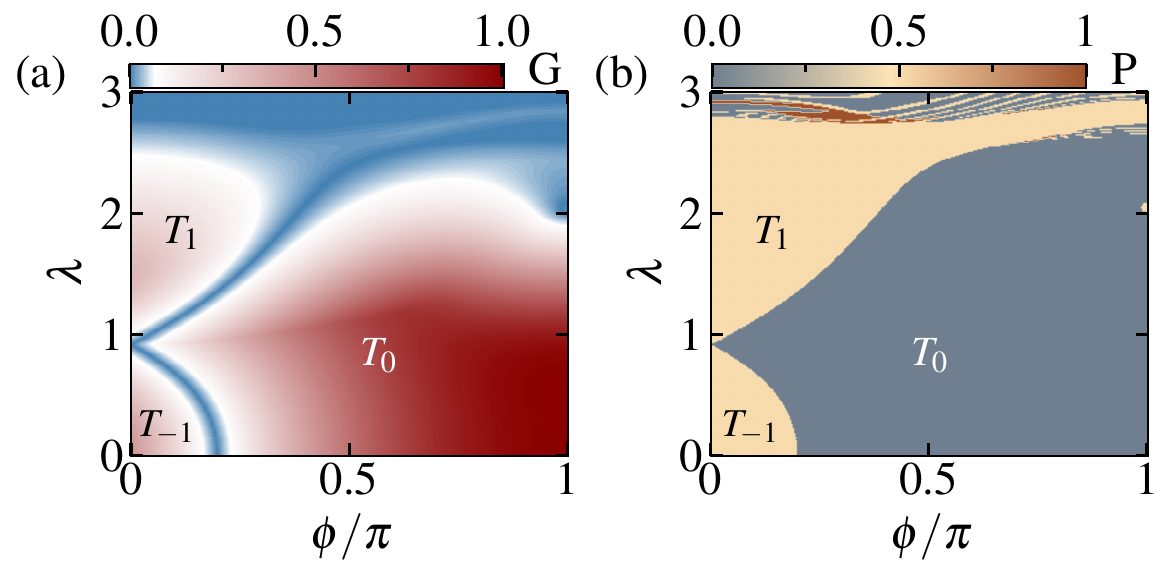}
\caption{(a) The gap $G$ as a function of $\lambda$ and  $\phi$ at fixed values of hopping parameters $t_p = t_2 =0.75,  t_1 = 1$. Blue regions denote gapless regions, whereas the reddish-white regions represent gapped topological (T$_1$, T$_-1$) and trivial (T$_0$) phases. (b) The polarization $P$ as a function of  $\lambda$ and $\phi$ to identify the topological and the trivial phases shown in (a). 
For all the plots, the phase of onsite disorder $\theta$ is $0.135\pi$, and the system size $L$ is $2584$. 
}
\label{fig:figure 2}
\end{figure}

\begin{figure*}[!t]
\centering
\includegraphics[width=2.0\columnwidth]{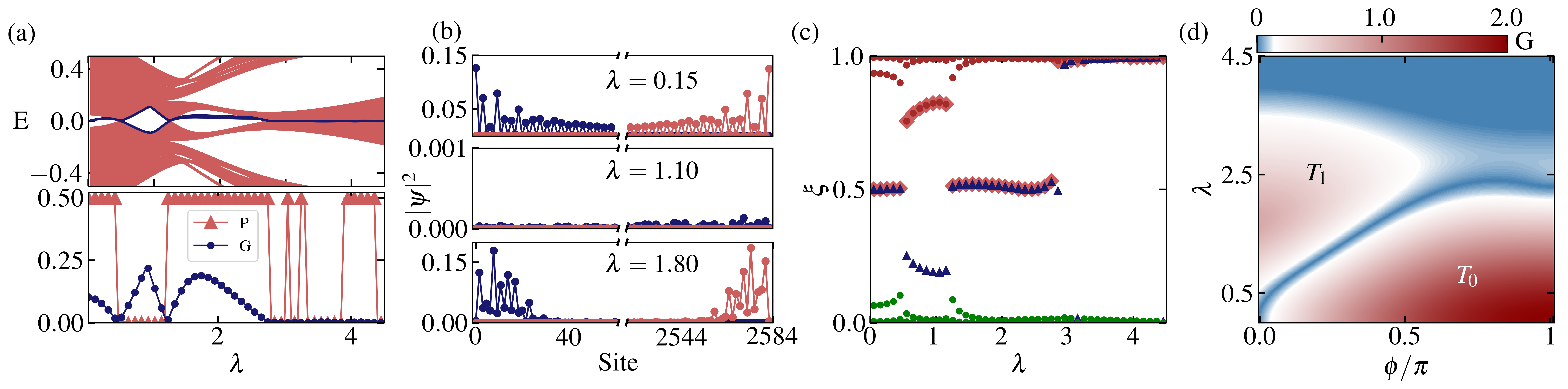}
\caption{
(a) The real-space energy spectrum under OBC as a function of disorder strength ($\lambda$), with fixed hopping parameters $t_p = t_2 = 0.75$, $t_1 = 1$, for $\phi = 0.15\pi$ (upper panel).  In the lower panel, the polarization ($P$) and the energy gap ($G$) as functions of $\lambda$, computed under PBC, using the same parameters as in the upper panel.  (b) The occupation probability of the mid-gap states $|\psi|^2$ is plotted for $\lambda = 0.15,~1.10,~\text{and}~1.80$ under OBC in the upper, middle, and lower panels, respectively. (c) Entanglement spectrum plotted as a function of $\lambda$.  (d) Gap $G$ is plotted as a function of  $\lambda$ and $\phi$ for fixed value of hopping parameters $t_p = t_1 = t_2 = 1$. Blue regions denote gapless regions, whereas reddish-white regions represent gapped topological (T$_1$) and trivial (T$_0$) phases.  In all the panels, the phase of onsite disorder is fixed at $\theta = 0.135\pi$, and the system size is $L = 2584$.
}
\label{fig:figure 3}
\end{figure*}

\paragraph*{Results.-}
The model shown in Eq.~\ref{eq:ham}, in the limit of $t_2=0$ and $\lambda=0$, reduces to the well known SSH chain~\cite{ssh_model} exhibiting a topological phase transition by varying $t_1~(t_p)$ while keeping $ t_p~(t_1)$ fixed.
However, for finite $t_2$ and in the absence of flux ($\phi=0$), the model maps to a two-leg ladder with opposite hopping dimerization patterns along the legs. This system is known to exhibit a topological phase for a suitable choice of $t_p$ and $t_2$, which is discussed in Ref.~\cite{padhan_ladder} by some of us. Keeping this scenario in mind, in this study, we fix $t_p = t_2=0.75$, so that the system is initially in the topological phase and analyze the effect of quasiperiodic disorder and flux on this topological phase.

We begin by highlighting our main findings, illustrated in the phase diagram of Fig.~\ref{fig:figure 2}. In Fig.~\ref{fig:figure 2}(a), we depict the gapped topological and trivial phases (reddish-white regions) separed from each other by the gapless regions (blue regions). The phase diagram is obtained by plotting the bulk gap $G = E_{\frac{L}{2}+1} - E_{\frac{L}{2}}$ computed under periodic boundary conditions (PBC) as a function of $\phi$ and $\lambda$. This shows that when $\phi=0$ and $\lambda=0$ (at the origin in Fig.~\ref{fig:figure 2}(a)), the system is in the gapped topological phase (T$_{-1}$) due to the choice of the hopping strengths (see Ref.~\cite{padhan_ladder}). Starting from this topological phase, when $\lambda$ is increased while keeping $\phi=0$ (along the y-axis of Fig.~\ref{fig:figure 2}(a)), the gap tends to decrease and completely vanishes  (a situation true for any topological phase~\cite{hayward_rice_mele, Nakajima_disorder, topo_anderson_pump, loc_topology_r_citro, disorder_pump_hoti,pumping_photonics}) at a critical $\lambda_c\sim0.92$. Interestingly, however, the gap reopens again with an increase in $\lambda$, leading to the emergence of another topological phase (T$_{-1}$) possessing different topological character from the former. 
This results in a phase transition between the two topological phases (T$_{1}$ and T$_{-1}$) through a gap-closing critical point. Remarkably, when $\phi$ is introduced in the system, another gapped phase emerges between T$_{1}$ and T$_{-1}$ phases, which we identify as a trivial gapped phase (T$_0$) that is separated from the topological phases by gapless lines.

To illustrate this behavior, we plot the single-particle energy spectrum as a function of $\lambda$ obtained under open boundary conditions (OBC) in the upper panel of Fig.~\ref{fig:figure 3}(a) for a cut through the phase diagram of Fig.~\ref{fig:figure 2}(a) at $\phi=0.15\pi$. 
At $\lambda=0$, there appears a gap in the middle of the spectrum which hosts two mid-gap states. 
This is the first indication of the topological nature of the system under OBC. 
As $\lambda$ increases, a gap-closing point appears in the spectrum at $\lambda \sim 0.5$, and edge states merge into the bulk (Fig.~\ref{fig:figure 3}(a)). 
However, the gap immediately reopens without any edge states, until a sufficiently strong disorder causes another gap closing scenario at $\lambda\sim 1.2$. 
Following this, the gap reopens once more, leading to the emergence of another set of mid-gap states, which persist until a final gap closing at $\lambda \sim 2.8$, where they again merge with the bulk.

To determine whether the blue-marked states are edge states or not, we analyze the onsite particle density, ($|\psi|^2$) corresponding to these states as a function of site index corresponding to three different cuts in Fig.~\ref{fig:figure 3}(a), i.e., $\lambda=0.15,~1.10 ~\text{and}~ 1.8$ shown as upper, middle, and lower panels in Fig.~\ref{fig:figure 3}(b), respectively. 
For $\lambda=0.15~\text{and}~1.80$, the upper and lower panels reveal finite probability density at the edges while being zero in the bulk, confirming the presence of edge states and indicating a topological nature. 
In contrast, the middle panel shows  uniform density throughout the bulk, ruling out any topological character of the system in this regime.
Note that the edge states are sensitive to the phase $\theta$, causing their point of merging into the bulk to vary across different disorder realizations~\cite{Phase}.

To confirm the nature of the gapped phases shown in the upper panel of Fig.~\ref{fig:figure 3}(a), we compute the real-space polarization ~\cite{resta_pol, hayward_rice_mele},
$P = \frac{Qa}{2\pi} \text{Im} \ln \langle \Psi | e^{-\frac{2\pi i}{N} X} | \Psi \rangle~\text{mod}~Qa$,
which acts as a topological invariant for distinguishing the topological and trivial phases in systems with disorder. 
The polarization $P$ is defined modulo $Qa$, where $Q$ denotes the charge and $a$ is the lattice spacing (taken to be unity in our study).
Here, $|\Psi\rangle$ is the many-body ground state at half-filling that is constructed from the single-particle states, and
\begin{align}
X = &\sum_{j \in\text{ even}} j n_{a,j} + \sum_{j \in\text{ odd}} (j+1) n_{a,j}\nonumber\\
&+ \sum_{j\in \text{ even}} (j+1) n_{b,j} + \sum_{j \in\text{ odd}} j n_{b,j},
\label{eq:pos_operator}
\end{align}
defines the total position operator~\cite{resta_pol}, where $N=L/2$. 
The topological (trivial) phase is identified from the value of $P=0.5 ~(0)$.
We plot $P$ (red triangles) as a function of $\lambda$ in the lower panel of Fig.~\ref{fig:figure 3}(a) for $\phi=0.15\pi$.
It can be seen that $P=0.5$ in the first gapped region, $0$ in the second, and $0.5$ in the third. Beyond this point, as the system becomes gapless, $P$ takes arbitrary values as expected.
For comparison, we show the gap $G$ (blue circles) as a function of $\lambda$.

To substantiate the reentrant topological phase transition, we examine the single-particle entanglement spectrum (ES)~\cite{Hughes1}, where the ground state under PBC is a single Slater determinant of half-filled single-particle states. For a spatial bipartition into subsystems A and B, the reduced density matrix is \( \rho_A = K e^{-H_{\text{ent}}} \), with the entanglement Hamiltonian \( H_{\text{ent}} = \sum_{aa'} h_{aa'} c_a^\dagger c_{a'} \), and where the indices $a,a'$ label states within A and h is related to the correlation matrix C by \( [C]_{aa'} = \text{Tr}(\rho_A c_a^\dagger c_{a'}) = [1/(e^h + 1)]_{aa'} \)~\cite{Hughes3, Hughes4}. The eigenvalues \( \{\xi_m\} \) of this $C$ (which lie between $0$ and $1$) represent the single-particle entanglement spectra.  and the entanglement entropy is given by \( S_A = -\sum_i [\xi_m \ln \xi_m + (1 - \xi_m) \ln (1 - \xi_m)] \).
In the topological phase, the entanglement spectrum features a robust midgap value at $\xi_m \approx 0.5$, yielding maximal entanglement that cannot be removed by any adiabatic deformation~\cite{Hughes1, Hughes2}. As shown in Fig. 3(d), for \( \lambda \approx 0 \) to $0.5$ and $1.2$ to $2.0$, the ES displays degenerate mid-gap \( \xi_m = 0.5 \) modes, indicating two non-trivial topological phases. However, in the intermediate range (\( \lambda \approx 0.5 \) to $1.2$), \( \xi_m \) gets away from 0.5 and collapses towards  $0$ or $1$, indicating a trivial phase. These observations corroborate the reentrant topological transitions inferred from both edge‐state spectra and polarization calculation, as shown in Fig.~\ref{fig:figure 3}(a). Beyond \( \lambda \approx 2.8 \), the gap closes, the ES loses the structure, and mid-gap modes vanish.

This analysis demonstrates a topological to trivial and then to a topological phase transition, which we call a re-entrant topological phase transition as a function of $\lambda$. This re-entrant phenomenon can be attributed to the nature of the onsite potential and the flux in the system. 
In the absence of any flux and disorder, the ladder can be viewed as two SSH chains of either $t_p$-$t_1$ type dimerization or $t_p$-$t_2$ type dimerization.  For the choice of hopping strengths in this case, when $\lambda=\phi=0$, the system realizes a topological phase (T$_{-1}$) whose topology originates from the $t_p$–$t_1$ type dimerization.  However, as $\lambda$ becomes finite, the effect of hopping dimerization in this configuration starts to decrease, and when $\lambda=\lambda_c\gtrsim 0.92$, the dimerization pattern changes to $t_p$–$t_2$ type, leading to another topological phase (T$_1$). This renormalization of the dimerization pattern happens due to the quasiperiodically modulated staggered onsite potential $\lambda$. On the other hand, when $\phi$ is increased starting from these topological phases, a transition to the trivial phase occurs due to another case of hopping renormalization and a change in dimerization pattern to a trivial configuration due to the flux in the system. This leads to two different gap-closing transition lines in the phase diagram (see Fig.~\ref{fig:figure 2}).

The role of the specific choice of staggered strength and quasiperiodic nature of onsite potential and flux on the emergence of disorder induced topological phase ($T_1$) at strong disorder strength can also be understood by starting from a uniform ladder (with uniform hoppings $t_1 = t_2 = t_p = 1$ and zero flux) which is initially a gapless liquid. Starting from this gapless liquid, increasing the strength of the quasiperiodic disorder alone induces a gapped topological phase as shown in Fig.~\ref{fig:figure 3}(d). 
Additionally, when a staggered flux is first applied to the clean uniform ladder, the initial gapless liquid becomes a trivial insulator; subsequently, increasing quasiperiodic disorder drives a trivial-to-topological phase transition. In this way, rung-staggered quasiperiodic disorder not only mediates the reentrant topological phase transition, requiring an initial topological phase, but can also counterintuitively induce a topological phase from an initially gapless or trivial phase. 
Note that in the uniform flux case, a reentrant topological phase transition does not occur (not shown); however, the behavior at $\phi=0$ and $\phi=\pi$ remains consistent with staggered flux cases.

In the following, we show that the two topological phases emerging in this scenario are fundamentally distinct, for which, we rely on one of the most fundamental and experimentally accessible mechanisms known as Thouless charge pumping ~\cite{Thouless1983, Nakajima_disorder, Liu2025, pumping_1d}.

\paragraph*{Thouless Charge Pumping.-}
\begin{figure}[t]
\centering
\includegraphics[width=1\columnwidth]{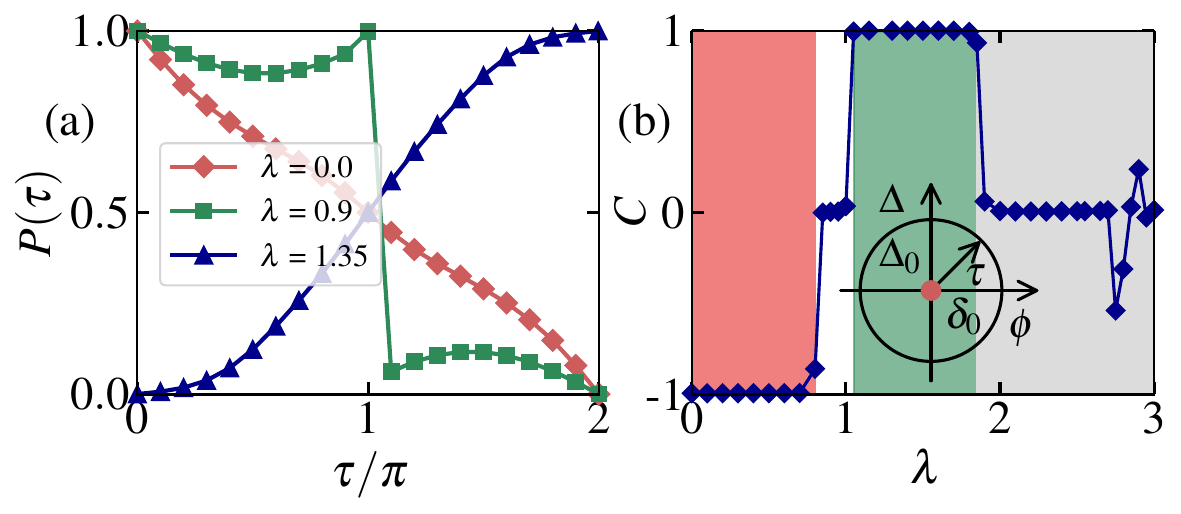}
\caption{(a) The evolution of polarization ($P$) as a function of $\tau$ at a cut through $\phi=0.2\pi$ of the phase diagram shown in Fig.~\ref{fig:figure 2} for three different values of $\lambda$. i.e. $0.0$ (red diamonds), $0.9$ (green squares), and $1.35$ (blue triangles). The pumping parameters are set as $\phi_{0}=0.2\pi, \delta_0=0.15\pi~~ \text{and},~\Delta_0=0.2$. The hopping amplitudes are set as $t_p=0.75$, $t_1=1.0$, and $t_2=0.75$ as consistent with the phase diagram shown in Fig.~\ref{fig:figure 2} (b) $C$ as a function of $\lambda$ for the same pumping parameters $\phi_{0}=0.2\pi, \delta_0=0.15\pi~~ \text{and},~\Delta_0=0.2$ for a system of size $L=2584$.}
\label{fig: figure 4}
\end{figure}
Thouless charge pumps (TCP) are the quantum analog of the Archimedes screw pump, which facilitates quantized particle transport through adiabatic and periodic modulation of some system parameters.
This phenomenon links quantum transport to topology, as the number of particles pumped over a full cycle is quantized and tied to a topological invariant, i.e., the Chern number of the system~\cite{pumping_quasicrystal,virtual_chern_num}. 
For quantized pumping to occur, the system must maintain an open bulk gap while undergoing a cyclic evolution that connects topological and trivial phases, enclosing the critical point of the topological phase transition in the parameter space. 
TCP has traditionally been studied in the context of the Rice-Mele model~\cite{rice_mele_model, Asboth2016_ssh}, which introduces a staggered potential to preserve the gap during the pumping process. 

In our case, introducing such staggered terms to our model leads to a disordered RM model:
\begin{equation}
H_{\text{DRM}} = H + \sum_{j} \Delta (-1)^j n_{a,j} + \sum_{j} \Delta (-1)^{j+1} n_{b,j},
\label{eq:drm_hamiltonian}
\end{equation}

In the clean limit ($\lambda=0$), with vanishing $t_2$, $H_{\text{DRM}}$ reduces to the original Rice-Mele model~\cite{rice_mele_model, Asboth2016_ssh}, where quantized Thouless charge pumping is well established~\cite{Lohse2016,Takahashi2016pumping,Hayward2018,monika_review,bound_pump1,bound_pump3,juliafare_quantum,Kuno2017,bertok_pump,mondal_phonon,spin_pumping,pumping_quasicrystal,Taddia2017,pumping_1d,hubbarad_thouless_pump,qubit_pumping,padhan_ladder,rajashri_ladder,seba_nphy_pumming,seba_nature_pumping,pumping_esslinger,pump_reversal_esslinger}. Recent experiments also have highlighted significant interest in Thouless charge pumping in disordered systems, particularly the interplay and competition between topology and quasiperiodic disorder in Thouless pumps~\cite{Nakajima_disorder, Liu2025}.

We demonstrate the TCP by adiabatically modulating the flux and staggered potential as $\phi = \phi_0 + \delta_0 \cos(\tau)$ and $\Delta = \Delta_0 \sin(\tau)$ respectively,  with $\tau$ as the pumping parameter, forming a closed loop in parameter space (See the inset in Fig.~\ref{fig: figure 4}(b)). 
Fig.~\ref{fig: figure 4}(a) shows the polarization $P$ versus $\tau$ for three disorder strengths: $\lambda=0.0$ (red diamonds), $0.9$ (green squares) and $1.35$ (blue triangles).
The hopping parameters are fixed at $t_p=t_2=0.75$, and $t_1=1.0$, while the pumping parameters are set to be $\phi_{0}=0.2\pi, ~\delta_0=0.15\pi~~ \text{and},~\Delta_0=0.2$.
From the evolution of polarization, we calculate the amount of charge pumped per cycle, which is given by
$Q = \int_0^{2\pi} d\tau ~\partial_\tau P(\tau)$.
At $\lambda=0$, $P$ changes smoothly from $1$ to $0$, corresponding to a quantised charge transfer $|Q|=1$. For $\lambda=0.9$, a discontinuity appears, marking the breakdown of pumping. Remarkably, at $\lambda=1.35$, $P$ again evolves smoothly but in the opposite direction ($0$ to $1$), yielding $|Q|=1$ again. This clearly marks a reversal phenomenon in the TCP. 

The above exercise also leads to another invariant quantity which identifies topological phases in systems with disorder called the local Chern marker (LCM)~\cite{lcm_resta}. 
The LCM is computed in the position basis as $C=\sum_iC_i$, with
\begin{equation}
    C_i = \frac{1}{\pi}\sum_{n=0}^{N_\tau -1}\text{Im}\langle i|X_e^\dagger\mathcal{P}(\tau_n)X_e\mathcal{Q}(\tau_n)\mathcal{P}(\tau_{n+1})\mathcal{P}(\tau_n)|i\rangle.
\end{equation}
Here, $\mathcal{P}$ and $\mathcal{Q}=\mathbb{1}-\mathcal{P}$ represent the projection operators into the occupied and unoccupied eigenstates of $H_{\text{DRM}}$, respectively, and $|i\rangle$ represents the single particle basis.
$X_e=\text{exp}(-X)$ defines the exponentiated position operator. 
The parameter $\tau$ is sampled over the interval [$0,~2\pi$] with a uniform discretization step $d\tau=\tau_{n+1}-\tau_n$, using $N_\tau$ total points.
The value of $C$ remains quantized as long as the bulk energy gap is finite during the pumping cycle.
To analyze this behavior, we compute $C$ for all values of $\lambda$ following different trajectories by appropriately modulating $\tau$, and by fixing the other parameters as $t_p = 0.75, t_1 = 1, t_2 = 0.75$, $\phi_0 = 0.2 \pi$, $\delta_0=0.15\pi$ and $\Delta_0=0.2$. 
Fig.~\ref{fig: figure 4}(b) depicts the red and green shaded regions with $C = 1$ and $-1$ indicating distinct topological phases at weak and strong disorder, where the pumping path encloses the topological phase transition point. 
The white region has $C = 0$, as the loop lies entirely within the trivial phase. 
In the grey region ($1.9 \lesssim \lambda \lesssim 2.7$), the loop stays completely within a topological phase without crossing a transition, giving $C = 0$; beyond this, the system becomes gapless and $C$ is no longer quantized.
We label the gapped phases as T$_{-1}$, T$_0$, and T$_1$ based on these distinct values of $C$ (see Fig.~\ref{fig:figure 2}). 
The change of $C$ from $-1$ to $1$ via $0$ with increasing $\lambda$ captures the reentrant topological transitions (compare with Fig.~\ref{fig:figure 2}) and also reflects a reversal in the direction of TCP.

\paragraph*{Conclusion.-}

We demonstrate that a two-leg flux ladder with rung-staggered quasiperiodic onsite disorder exhibits a remarkable disorder-induced reentrant topological phase transition, where the two topological phases appearing in the system possess opposite Chern markers, which in turn realizes reversal of Thouless charge pumping direction. The findings reveal a non-standard phenomenon of disorder-induced topological phase and phase transitions in one dimension. Such findings are not only  hold fundamental significances but are may have important implications in the context of technological applications involving robust topological transport~\cite{Citro2023}. 
Moreover, the findings provide directions for further investigation of disorder and flux effects on interaction induced topology~\cite{mondalv1v2, harsh, parida2025, padhan_ladder, rajashri_ladder}. The recent progress in experimental techniques for the quantum simulation of quasiperiodic lattices~\cite{BRYCE1, B6, B8, B9, B10, B11, B12, Rechtsman_2023_reentarnt, Li2023, Zhang_PrB_circuit} including flux lattices~\cite{ STAGG2, Exp1, Exp2, Exp3, BRYCE1, Exp4, BRYCE2} using ultracold atoms in optical lattices~\cite{BRYCE1, STAGG2,  Nakajima_disorder} may provide path for the observation of the results we have obtained.

\paragraph{Acknowledgement.-} We thank Adhip Agarwala for useful discussions. T.M. acknowledges support from Science and Engineering Research Board (SERB), Govt. of India, through project No. MTR/2022/000382 and STR/2022/000023.

\bibliography{ref}

\begin{thebibliography}{81}
\bibitem [{Pha()}]{Phase}%
  \BibitemOpen
  \bibfield  {title} {\enquote {\bibinfo {title} {Here, we choose the value of $\theta=0.135\pi$ for enhanced visibility of the edge states in the bulk gap.}}\ }\href@noop {} {\ }\BibitemShut {NoStop}%
\end{thebibliography}

\end{document}